\documentclass{osa-article}

\journal{osajournal}

\articletype{Research Article}

\begin{document}

\title{Triply-Resonant Sum Frequency Conversion with Gallium Phosphide Ring Resonators} 

\author{Alan D. Logan,\authormark{1,*} Shivangi Shree,\authormark{2} Srivatsa Chakravarthi,\authormark{2} Nicholas Yama,\authormark{1} Christian Pederson,\authormark{2} Karine Hestroffer,\authormark{3} Fariba Hatami,\authormark{3} and Kai-Mei C. Fu\authormark{1,2,4}}

\address{\authormark{1}Department of Electrical and Computing Engineering, University of Washington, Seattle WA 98195, USA\\
\authormark{2} Department of Physics, University of Washington, Seattle WA 98195, USA\\
\authormark{3}Department of Physics, Humboldt-Universitat zu Berlin, 12489 Berlin, Germany\\
\authormark{4} Physical Sciences Division, Pacific Northwest National Laboratory, Richland, Washington 99352, USA}
\email{\authormark{*}adlogan@uw.edu} 



\begin{abstract*}
We demonstrate quasi-phase matched, triply-resonant sum frequency conversion in 10.6-\textmu m-diameter integrated gallium phosphide ring resonators. A small-signal, waveguide-to-waveguide power conversion efficiency of 8\%/mW is measured for conversion from telecom (1536\,nm) and near infrared (1117\,nm) to visible (647\,nm) wavelengths with an absolute power conversion efficiency of 6.3\% measured at saturation pump power. For the complementary difference frequency generation process, a single photon conversion efficiency of 7.2\%/mW from visible to telecom is projected for resonators with optimized coupling. Efficient conversion from visible to telecom will facilitate long-distance transmission of spin-entangled photons from solid-state emitters such as the diamond NV center, allowing long-distance entanglement for quantum networks.  %
\end{abstract*}

\section{Introduction}

Difference and sum frequency conversion in $\chi^{(2)}$ materials provide an efficient and general means to access new wavelengths with established sources.
Conversion efficiency in these processes scales with field intensity, motivating resonant geometries to both increase per-unit power conversion efficiency and decrease the interaction volume and thus footprint. Resonant geometries are of particular interest for applications in quantum information in which devices must operate at the single photon level. Single photon frequency conversion processes, also called quantum frequency conversion (QFC), will enable a wider range of qubits to access telecommunication wavelengths, enabling long-distance entanglement for fiber-based quantum repeater networks~\cite{kumar1990qfc,bock2018high,van2020long,zaske2012visible,tchebotareva2019entanglement, walker2018long,dreau2018nvqfc}. Further, fine control of the converted photon wavelength could be used to erase inhomogeneities in emission wavelength, permitting high-fidelity remote entanglement of inhomogeneous qubits by two-photon interference.~\cite{ates2012two,weber2019twophoton,levonian2022distinguishable}.
QFC may also be useful as an upconversion mechanism for efficient room-temperature photon detection~\cite{albota2004bulkqfc,rutz2017quantum} and could provide a bridge between disparate qubit platforms~\cite{allgaier2017conversion}.

QFC based on a material's second order nonlinear susceptibility $\mathrm{\chi^{(2)}}$ utilizes a strong coherent pump to enable interactions between two single-photon modes, so it is restricted to sum- and difference-frequency generation (SFG and DFG) processes with three distinct wavelengths. Single-photon SFG/DFG has been demonstrated in macroscopic optical cavities~\cite{albota2004bulkqfc,samblowski2014conversion} and PPLN waveguides~\cite{guerreiro2013interaction,pelc2011long,dreau2018nvqfc}, with internal maximum photon conversion efficiencies as high as 86\% for pump powers of hundreds of mW. 
In nanophotonic resonators, triply resonant QFC processes have been demonstrated using two detuned inputs on a single resonance~\cite{stolk2022telecom,wang2021degenerate} or adjacent azimuthal resonances of the same transverse mode~\cite{ye2020triple}. In both cases, two of the three wavelengths involved in the conversion process are restricted to a range of a few nanometers. 
However, implementing QFC interfaces for specific quantum emitters, e.g. the nitrogen-vacancy~\cite{dreau2018quantum,ikuta2014frequency} or silicon-vacancy~\cite{huang2021siv,bersin2021siv} in diamond, will typically require long jump frequency changes with three resonances at widely disparate wavelengths. 

As a step toward low power visible-to-telecom QFC for diamond NV centers, we demonstrate triple-resonant, quasi-phase matched SFG from the telecom C-band to visible red in gallium phosphide (GaP) ring resonators on silicon oxide. GaP is transparent into the visible band and provides both high refractive index (3.31 at 637nm)\cite{bond1965measurement} and high nonlinear susceptibility $\chi^{(2)} \sim \mathrm{110 pm/V}$ \cite{dal1996density}, providing a versatile nonlinear photonic layer compatible with a wide variety of substrates \cite{rivoire2011second,barclay2009hybrid,schneider2018gallium,englund2010deterministic}. We present the ring resonator design, an efficient method to identify triply-resonant phase matched devices, and the overall performance for the highest performing devices. We measure a small-signal, SFG (telecom-to-visible) waveguide-to-waveguide power conversion efficiency of 8\%/mW. 
For conversion from near-infrared to visible, a saturated absolute power conversion efficiency of 6.3\% was measured using a pulsed 1.8\,mW telecom laser as the pump.
The projected SFG (telecom-to-visible) power conversion efficiency of the current devices with optimized waveguide coupling is $\mathrm{\sim 17\%/mW}$, which corresponds to a DFG (visible-to-telecom) photon conversion efficiency of $\mathrm{\sim 7.2\%/mW}$.

\section{Model and design}
The GaP ring resonators were designed to maximize waveguide-to-waveguide on-chip conversion efficiency.
For a process that converts visible photons in the $\omega_3$ input mode to telecom photons in the $\omega_1$ output mode using a pump at $\omega_2 = |\omega_3 - \omega_1|$, the theoretical power conversion efficiency $\eta$ in the low-power limit is \cite{burgess2009difference}: 

\begin{equation} \label{eq:conv_eff} 
    \frac{\eta}{P_{2,in}} = \frac{P_{1,out}}{P_{2,in} P_{3,in}} = 64 \left| \beta \right|^{2} \frac{\omega_{1}}{\omega_{2}\omega_{3}} \frac{Q_1^2 Q_2^2 Q_3^2}{Q_{c1} Q_{c2} Q_{c3}}
\end{equation}
where $Q_n$ and $Q_{cn}$ are the total and coupling quality factors of mode $n$, and $P_n$ is the input/output power for each mode.
Small-signal or normalized conversion efficiency is defined as $\eta_{ss} = P_{1,out}/P_{2,in} P_{3,in}$.
Because the quality factors are primarily limited by material properties and fabrication, we focus on maximizing the nonlinear mode overlap $\beta$, which encompasses the effects of mode field distribution and polarization, phase matching, and mode volume:

\begin{equation} \label{eq:beta_xtal} 
    \beta = \frac{1}{4\sqrt{\varepsilon_{0}}} \frac{\int \sum_{i,j,k} \chi^{(2)}_{ijk} E^{*}_{3i} \left( E_{1j} E_{2k} + E_{2j} E_{1k} \right)  \, dV}{\sqrt[]{\int \epsilon_1 |E_1|^2 \, dV} ~\sqrt[]{\int \epsilon_2 |E_2|^2 \, dV} ~ \sqrt[]{\int \epsilon_3 |E_3|^2 \, dV}}.
\end{equation}
$E_n$ and $\epsilon_n$ are the electric field distribution and relative permittivity for mode $n$, and $i$, $j$, and $k$ are crystallographic directions of the nonlinear medium.
The ring resonators were designed for 430\,nm thick GaP on silicon oxide.
Due to the zincblende crystal symmetry of GaP, $\chi^{(2)} \neq 0$ for $i \neq j \neq k$, so nonlinear interactions require mutually perpendicular field components in each mode.
For a radially symmetrical structure in [1 0 0]-normal GaP, the nonlinear overlap $\beta$ becomes:

\begin{align} \label{eq:beta_ring}
\begin{split}
\beta &= \frac{ \left| \chi^{(2)} \right|}{4\sqrt{\varepsilon_{0}}} \frac{\int_{NL}  \left[ (B_1 - iB_2) e^{i(M + 2)\theta} + (B_1 + iB_2) e^{i(M - 2)\theta} \right] \, dV}{\sqrt[]{\int \epsilon_1 |E_1|^2 \, dV} ~\sqrt[]{\int \epsilon_2 |E_2|^2 \, dV} ~ \sqrt[]{\int \epsilon_3 |E_3|^2 \, dV}}  \\
M &= m_1 + m_2 - m_3 \\
B_{1} &=  E_{1z} (E_{2r} E_{3\theta}^{*} + E_{3r}^{*} E_{2\theta}) + E_{2z} (E_{1r} E_{3\theta}^{*} + E_{3r}^{*} E_{1\theta}) + E_{3z}^{*} (E_{1r} E_{2\theta} + E_{2r} E_{1\theta}) \\
B_{2} &= E_{1z} (E_{2r} E_{3r}^{*} - E_{2\theta} E_{3\theta}^{*}) + E_{2z} (E_{1r} E_{3r}^{*} - E_{1\theta} E_{3\theta}^{*}) + E_{3z}^{*} (E_{1r} E_{2r} - E_{1\theta} E_{2\theta})
\end{split}
\end{align}
where $E_{ir}$, $E_{iz}$, and $E_{i\theta}$ are the radial, vertical, and azimuthal field components and $m_{i}$ is the azimuthal mode number of each mode.

The nonlinear overlap exhibits a sinc-like dependence on the quasi-phase matching (QPM) condition $M = \pm 2$. 
The overlap is greatly diminished for single- or double-resonant processes, and vanishes entirely for unmatched triple resonance ($M \neq \pm 2$).
Consequently, when the designed set of modes are tuned to triple resonance, it is unlikely that any other frequency conversion process with the same input wavelengths, such as second harmonic generation, will be significantly enhanced by the resonator.
Each term in the overlap is a product of one vertical and two in-plane field components, which in practice restricts mode selection to one TM and two TE modes. 
The difference in interactions between radial and azimuthal field components in the $M = +2$ and $M = -2$ cases biases the overlap integral toward the inside and outside edge of the ring, respectively, which allows significant interactions between modes with even and odd symmetry \cite{thiel2022trimming}.

The ring resonator was designed using the Lumerical MODE bent-waveguide solver to simulate propagation constants and nonlinear overlap for telecom, NIR, and visible modes. 
Two designs were explored. In the first design, the ring width and radius were optimized to maximize $\beta$ while satisfying the $M = -2$ quasi-phase matching condition for $\mathrm{TE_{00}}$, $\mathrm{TE_{00}}$, and $\mathrm{TM_{03}}$ modes at 1550\,nm, 1081.4\,nm, and 637\,nm, respectively (Supplement 1). The final ring dimensions for this design are a radius of 5.3\,\textmu m, measured to the center of the ring, with ring width varied from 675-704\,nm in 1\,nm increments to compensate for fabrication variations.
The second design utilized $\mathrm{TE_{00}}$, $\mathrm{TE_{01}}$, and $\mathrm{TM_{04}}$ modes at 1550\,nm, 1081.4\,nm, and 637\,nm. This mode combination allowed more than a two-fold increase in simulated $\mathrm{|\beta^2|}$ compared to the first design in exchange for potentially reduced quality factors. The fabricated devices exhibited lower SFG efficiency compared to the first design (Supplement 1) and will not be discussed further here.

\begin{figure}[ht!]
\centering\includegraphics[width=\textwidth]{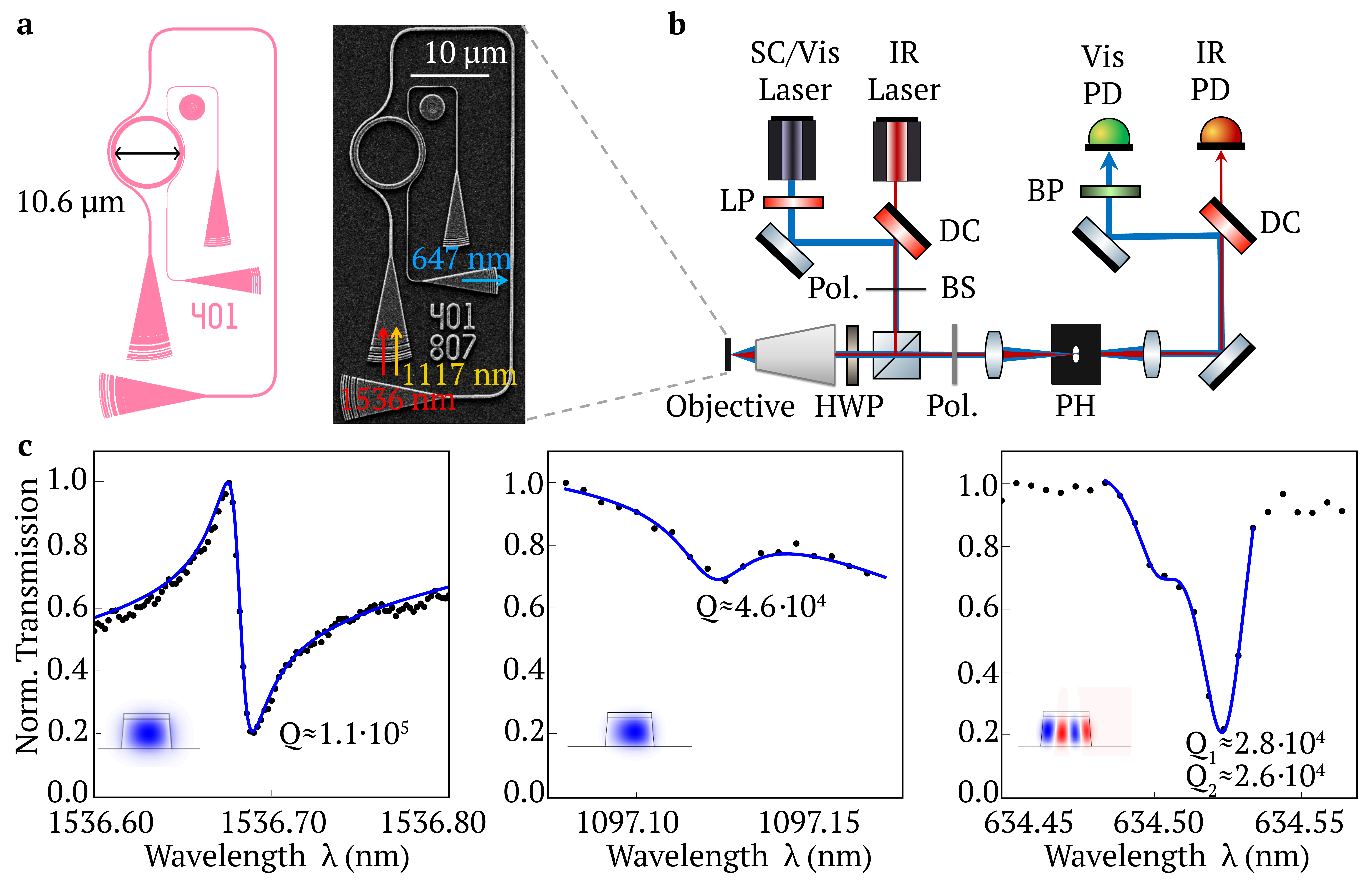}
\caption{(a) Device layout and SEM image of a GaP nonlinear ring resonator coupled to input/output waveguide circuits, one for 637-647\,nm visible light (blue) and one combining 1530-1565\,nm telecom (red) and 1080-1120\,nm NIR (yellow) light. Each coupling circuit includes two grating couplers to allow single-wavelength transmission measurements in addition to the SFG measurement marked on the SEM. (b) Free-space coupled microscope setup for device characterizations and SFG measurements. Cross-polarization and the pinhole (PH) are used to eliminate reflected/scattered input light. IR: infrared, SC: supercontinuum, PD: photodiode, BS: beamsplitter, DC: dichroic, HWP: half-wave plate, BP: band-pass filter, LP: long-pass filter. Input and output filter wavelengths were adjusted for different measurement wavelengths. (c) The transmission measurements from telecom, near-infrared and visible resonances in device SFG01, along with fitted Fano/Lorentzian curves. Background is approximated with linear functions. Ring cross-sections with vertical (visible) or radial (telecom and NIR) electric field distributions are inset for each mode. Transmission at each wavelength was measured using a tunable laser input and wavelength-appropriate photodiode. Due to difficulties with measuring and fitting the specific resonances used for triply-resonant SFG testing, the fits shown here use different instances of the same family of resonances.}
\label{fig:layout}
\end{figure}

From Eq.~\ref{eq:conv_eff}, the small-signal conversion efficiency is maximized when all three modes are critically coupled ($Q_c = 2Q$), while the quantum limit of the photon conversion efficiency can be increased to approach unity by overcoupling the input and output modes \cite{burgess2009difference}.
With a simulated nonlinear overlap of $|\beta|^2 = \mathrm{27.8\,J^{-1}}$ for the first ring design, a critically coupled device with a combined total quality factor of $(Q_1 Q_2 Q_3 = 10^{15})$ is simulated to provide a waveguide-to-waveguide photon conversion efficiency of $\mathrm{\sim 13.5\%/mW}$ between visible and telecom wavelengths.
Wrapped-waveguide coupling regions were designed to provide a range of target coupling quality factors for each mode (Supplement 1). 
As shown in Fig.~\ref{fig:layout}a, both infrared modes were coupled to a single waveguide, while the visible mode was coupled to a separate waveguide.

\section{Fabrication}

The photonic devices were fabricated in a 427\,nm-thick [1 0 0]-normal GaP membrane on 1\,{\textmu}m $\mathrm{SiO_{2}}$ on silicon.
The GaP membrane was grown by gas-source molecular-beam epitaxy on an intermediate 300\,nm $\mathrm{Al_{0.8}Ga_{0.2}P}$ sacrificial layer on a bulk GaP substrate.
The sacrificial layer was etched in 3:100 HF:$\mathrm{H_{2}O}$ to release the 2\,mm square GaP membrane.
The membrane was wet transferred to the oxide substrate \cite{logan2018shg}, which was cleaned with solvents, $\mathrm{O_2}$ plasma, and treated with HMDS adhesion-promoter.
The photonic circuit pattern was written by 100\,keV electron beam lithography in a $\mathrm{\sim 100\,nm}$ HSQ resist layer, which was developed with 25\% TMAH in water.
The resist pattern was transferred to the GaP layer by an inductively-coupled plasma reactive ion etch (ICP-RIE) using a $\mathrm{Cl/Ar/N_2}$ (1.0/6.0/3.0 sccm) plasma, which selectively stops at the oxide layer. 
After fabrication, the HSQ resist could not be removed without potentially damaging the GaP or the oxide substrate, so it was left as partial cladding.

\section{Passive device characterization}

The setup shown in Fig.~\ref{fig:layout}b was used for both passive and frequency-conversion device testing.
Passive device testing consisted of transmission measurements at each wavelength band, which were used to determine the quality factors of the three modes and the grating coupler efficiencies near the resonant wavelengths.
The telecom resonances were characterized by scanning an input laser from 1530\,nm to 1565\,nm and recording the device transmission on an infrared photodiode.
The measured total quality factors of the telecom $\mathrm{TE_{00}}$ mode ranged from $\mathrm{\sim 0.9-2.9 \times 10^5}$.
Visible and near-infrared resonances were characterized via transmission spectroscopy with a broadband supercontinuum laser input and grating spectrometer collection.
Devices with promising SFG performance were selected for further characterization using scanning narrowband sources (632-640\,nm for visible, 1070-1130\,nm for NIR).
In this subset of devices, visible and NIR quality factors of $\mathrm{\sim 2.5-3 \times 10^4}$ and $\mathrm{\sim 4.5-6 \times 10^4}$ were measured, respectively.
The transmission measurements for the most efficient SFG device (SFG01) are shown in Fig.~\ref{fig:layout}c, with telecom, NIR, and visible quality factors of $\mathrm{1.1 \times 10^5}$, $\mathrm{4.6 \times 10^4}$, and $\mathrm{2.8 \times 10^4}$, respectively.

\begin{figure}[hbt]
\centering\includegraphics[width=\textwidth]{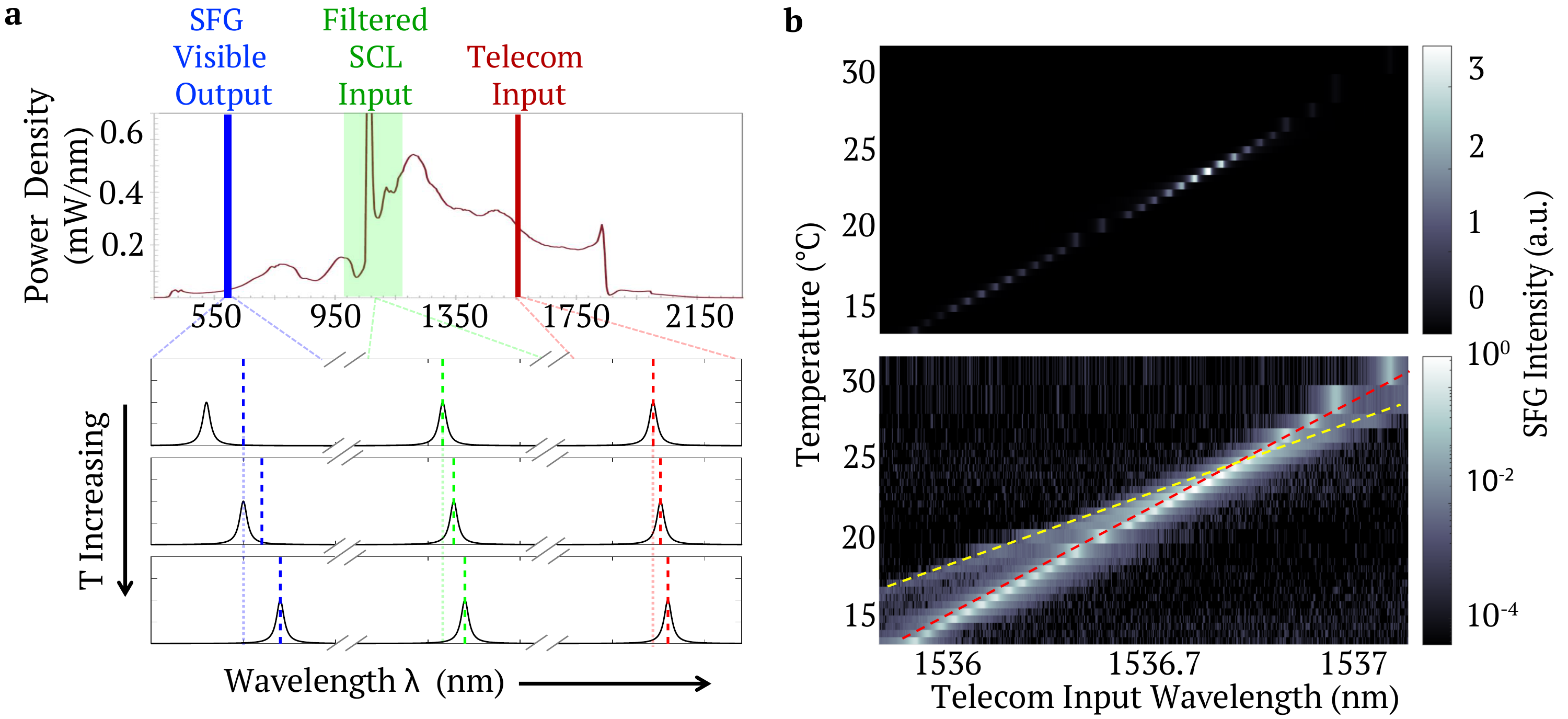}
\caption{(a) The power density of the Whitelase Micro supercontinuum  (SCL) laser, with the filtered wavelength band used as an input for SFG marked in green. The wavelength bands for the telecom tunable input laser (red) and visible output collection filter (blue) are also marked. Below is an illustration of the temperature-dependent SFG measurement used to identify triple-resonant SFG processes as described in the text. (b) Supercontinuum-pumped SFG measurements for device SFG01, showing dependence on device temperature and telecom-band input wavelength. Top is a linear-scale colormap and bottom is a logarithmic-scale colormap. The red line marks the telecom-band resonance, and the yellow line marks the input wavelengths that produce SFG light on resonance with the visible-band mode. Conversion efficiency is maximized when both conditions can be satisfied simultaneously (triple resonance).}
\label{fig:scl_sfg}
\end{figure} 

\section{Triply-resonant device identification}

Finding devices that could be tuned to quasi-phase matched triple resonance based on the passive transmission measurements proved difficult due to shallow transmission dips from non-critically coupled resonances in many devices, along with the highly multi-mode nature of the ring resonators in the visible band.
Instead, we identified triply-resonant devices using rapid measurements of frequency conversion performance.
As illustrated in Fig.~\ref{fig:scl_sfg}, frequency conversion processes that are close to triple resonance have greatly enhanced efficiency which depends strongly on the detuning between the converted light and the output resonance wavelength.
Consequently, a near triply-resonant device under resonant excitation of two input modes will produce converted light that is both particularly bright and highly sensitive to detuning determined by the device temperature.

Due to the availability of high-quality lasers in the telecom band and detectors in the visible band, we used telecom-to-visible SFG measurements to characterize frequency conversion performance. The designed complementary DFG process is expected to have similar photon conversion efficiency.
In order to efficiently explore SFG performance in a three-dimensional parameter space (two input wavelengths and device temperature), we use a broadband supercontinuum laser (SCL) as the pump-band input (filtered to 1000-1200\,nm), and combine it with a tunable laser in the telecom band (1530-1565\,nm) with approximately 300\,{\textmu}W of power in the waveguide.
Converted visible light is collected using a 775\,nm short pass filter, 637/22\,nm bandpass filter, and a spatial filter to remove both scattered supercontinuum light and second harmonic converted light from the telecom laser.
Some portion of the supercontinuum pump provides resonant excitation to every NIR-band mode.
Now, when the telecom laser scans over a resonance, visible light is produced by SFG processes within the ring, enhanced by the telecom and a NIR-band mode.
The observed intensity of the converted visible light is a function of the detuning of the constituent resonances. A phase-matched visible resonance will not contribute much additional enhancement if the SFG wavelength is substantially detuned.
The detuning can be modified by changing device temperature, but the SFG efficiency is not significantly affected until the SFG wavelength approaches the resonance.
At this point, SFG efficiency becomes highly sensitive to temperature and peaks when the converted light is on resonance.
Consequently, a strong temperature dependence in the intensity of the SCL-pumped SFG in a device indicates a phase-matched triple-resonant SFG process.

We performed supercontinuum-pumped SFG measurements on 123 ring resonators at temperatures from $\mathrm{16-40~^{\circ}C}$ in $\mathrm{4~^{\circ}C}$ increments. Most of the tested devices produced some measurable SFG light within this temperature range.
Eight resonators with bright, strongly temperature-dependent SFG were selected for further characterization and designated SFG01-SFG08.
Supercontinuum-pumped SFG measurements were repeated with a finer temperature sweep to find the optimal operating temperature for each device. 
As shown for device SFG01 in Fig.~\ref{fig:scl_sfg}b, the SFG intensity peak red-shifts with increasing temperature to follow the high-Q telecom resonance.
The peak intensity is maximized at a device temperature of $\mathrm{\sim 25^{\circ}C}$ at which the condition of triple resonance met. This condition is apparent on a logarithmic scale in which two SFG peaks can be observed at each temperature: the telecom-band resonance, and a second peak at the input wavelengths that produce SFG light on resonance with the visible-band mode.
Similar measurements for the other seven devices are presented in Supplement 1.

Spectra of the SFG output were taken to determine the visible SFG wavelength. The visible SFG wavelength combined with the measured telecom input resonance uniquely determines the pump frequency. 
As shown in Table~\ref{tab:sfg_temp}, the narrower ring resonators exhibited triple resonance with modes at $\mathrm{\sim 1550}$, $\mathrm{\sim 1107}$, and $\mathrm{\sim 647\,nm}$, and wider rings at $\mathrm{\sim 1536}$, $\mathrm{\sim 1117}$, and $\mathrm{\sim 647\,nm}$. 
The discrepancy between the designed and measured process wavelengths may be due to fabrication variations, particularly in the ring coupling regions and sidewall angles.

\begin{table*}[ht]
\vspace{2em}
\begin{center}
\footnotesize
\definecolor{aliceblue}{rgb}{0.94, 0.97, 1.0}
\definecolor{lightgray}{rgb}{0.9, 0.9, 0.9}
\begin{tabular}{|p{0.09\textwidth} | p{0.09\textwidth} | p{0.09\textwidth} | p{0.09\textwidth} | p{0.09\textwidth} | p{0.09\textwidth} | p{0.09\textwidth} |}
\hline
\rowcolor{aliceblue}
Device & Width (nm) & $\mathrm{Opt.~T}$ ($^{\circ}$C) & $\mathrm{Est.~\eta_{\gamma}}$ $\mathrm{(\%/mW)}$ & $\mathrm{Telecom~\lambda}$ (nm) & $\mathrm{SFG~\lambda}$ (nm) & $\mathrm{Pump~\lambda}$ (nm) \\\hline 
SFG01 	&  698  &  25 & 8.00 & 1536.7    & 646.82    &  1117.0 \\\hline 
SFG02   & 682   & 18    & $\mathrm{5.75^*}$     & 1552.5    & 646.51    & 1107.2\\\hline 
SFG03   & 684   & 25    & $\mathrm{\sim 5.7}$    & 1552.8    & 646.59    & 1107.9\\\hline 
SFG04 	& 682   & 22.5  & 4.38   & 1552.4    & 646.4     & 1107.6 \\\hline 
SFG05   & 699   & 22    & 1.77     & 1536.3    & 646.66    & 1116.7\\\hline 
SFG06   & 701   & 25    & 1.36  & 1537      & 647.66    & 1119.4\\\hline  
SFG07 	& 679   & 18.5  & 0.91 & 1550.9    &  645.9    & 1106.9\\\hline 
SFG08   & 696   & 31    & 0.38  & 1536.5    & 646.66    & 1116.6\\\hline 
\end{tabular}
\end{center}
\caption{Ring width, operating temperature, efficiency and estimated SFG process wavelengths for SFG processes in eight triply-resonant ring resonators, based on supercontinuum-pumped SFG measurements. The visible wavelength of the dominant SFG process was measured with a spectrometer, and the pump-band wavelength was calculated from the telecom input and SFG output wavelengths. Conversion efficiency was measured directly for device SFG01, and estimated for other devices based on a comparison to this value. The efficiency estimate for device SFG03 is less precise due to a damaged visible-band output grating coupler. \\ $\mathrm{^{*}}$ Narrow-band SFG measurements of device SFG02 found a significantly lower conversion efficiency (Supplement 1).}
\label{tab:sfg_temp}
\end{table*}

\section{Frequency conversion efficiency}

\begin{figure}[h!]
\centering\includegraphics[width=\textwidth]{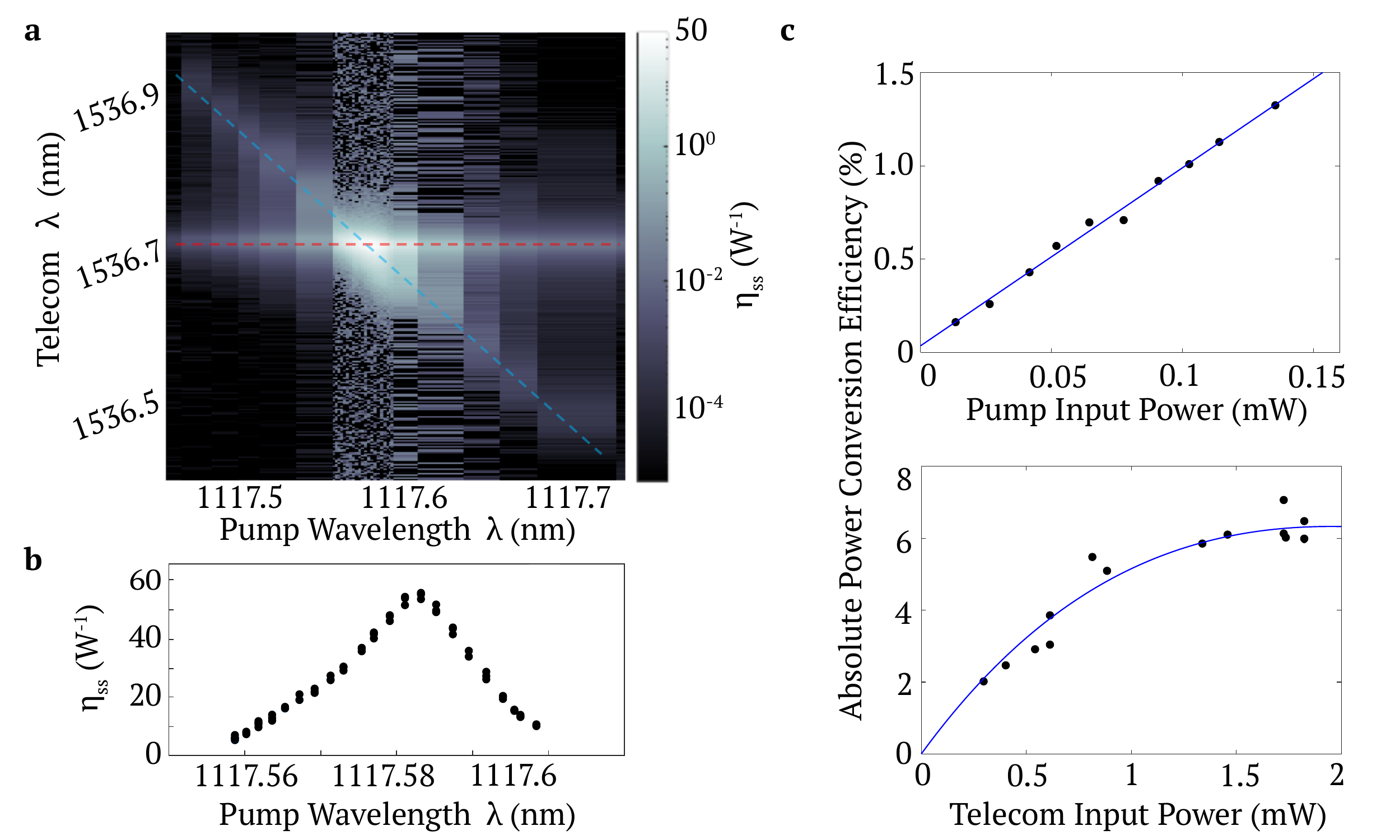}
\caption{ (a) Input wavelength dependence of small-signal SFG power conversion efficiency for device SFG01 at its optimal temperature ($\mathrm{25^{\circ}C}$), using CW telecom and pump inputs with average powers of 100\,\textmu W and 82\,\textmu W in the waveguide. The red line marks the SFG enhancement from resonant excitation of the telecom mode, and the blue line shows enhancement from to the visible mode applied to combinations of input wavelengths that produce an on-resonance SFG wavelength.
The collected SFG signal was attenuated to avoid detector saturation when the NIR wavelength approached resonance, resulting in exaggerated background noise from $\mathrm{\sim 1117.55-1117.65\,nm}$. (b) Detail of the small-signal conversion efficiency peak for device SFG01 around the NIR resonance. The linewidth of the efficiency peak corresponds to a pump mode quality factor of $\mathrm{5.7 \times 10^4}$, with a maximum conversion efficiency of $\mathrm{\sim 5\%/mW}$. (c) Power dependence of absolute conversion efficiency for device SFG01. The top plot shows conversion using low-power CW inputs, varying the NIR-band input power. The bottom plot shows conversion using a constant, low-power CW NIR input with a variable-power pulsed telecom input, which allows higher instantaneous excitation powers while avoiding the effects of resonator heating. }
\label{fig:sfg}
\end{figure}

While the broadband SCL pump allows efficient identification of triple-resonant devices, continuous and narrow-band pulsed inputs allow more precise control of the power used to excite a targeted frequency conversion process.
We measure the SFG small signal conversion efficiency of the brightest devices (SFG01 and SFG02) with a 1080-1120\,nm tunable CW laser as a pump in place of the SCL.
At the optimal temperature found with the SCL measurements, a 2D scan of both excitation lasers over their respective resonances was performed while monitoring the input power from each laser as shown in Fig.~\ref{fig:sfg}a,b. Similar to the temperature tuning data in Fig.~\ref{fig:scl_sfg}b, we again observe SFG enhancement from the visible (blue line) and telecom resonances (red line).
SFG measurements with the pump laser near its resonance required additional attenuation to avoid saturating the detector with the peak intensity. Rescaling the signal to compensate for the attention resulted in a heightened noise floor, hiding the enhancement from the pump-band resonance alone.
Waveguide-to-waveguide small signal conversion efficiency was calculated from measured input and output powers, grating efficiency, and microscope efficiency (Supplement 1).
For device SFG01, the peak small-signal efficiency found with this measurement was 4.9\%/mW.

Fig.~\ref{fig:sfg}c (top) gives the absolute power conversion efficiency for SFG01 in the small signal regime. A linear dependence on pump power is observed as expected. To obtain the maximum power conversion, the roles of the telecom and NIR laser were swapped due to the need for higher pump power which can be achieved with an erbium doped fiber amplifier added to the telecom input. Additionally, to minimize resonator heating and optical bistability, a telecom electro-optic modulator was used to pulse the telecom laser. The weaker NIR laser remained in CW mode. We note that the pulse lengths (100\,ns) and duty cycle (1\%) used are comparable to what would be utilized in single diamond defect applications. As shown in Fig.\ref{fig:sfg}c (bottom), an absolute power conversion efficiency of 6.3\% was measured for a peak telecom input power in the waveguide of 1.8\,mW. 
In devices targeting single photon conversion, pump depletion is not a factor and over coupling of the single photon modes will minimize back-conversion, allowing much higher maximum conversion efficiency.
Fitted curves for the CW and pulsed power dependence measurements yielded small-signal conversion efficiencies 9.5\%/mW and 8\%/mW, respectively. The variation in measured conversion efficiency may be due to slight variations in input laser alignment or gradual changes in the HSQ cladding~\cite{thiel2022trimming} altering the optimal temperature of the device.

The small-signal conversion efficiency found for device SFG02 with similar measurements was $\mathrm{\sim 0.37-0.59\%mW}$ (Supplement 1).
Estimates of the efficiency for devices SFG03-SFG08 were found by comparing the supercontinuum-pumped SFG intensities to device SFG01, and are given in Table~\ref{tab:sfg_temp}.

\section{Conclusion and outlook}

In triple resonant ring resonators, we demonstrate a waveguide-to-waveguide small signal power SFG conversion efficiency of 8\%/mW and measure a 6.3\% absolute power conversion efficiency.
At critical coupling, the expected SFG power conversion efficiency is $\mathrm{\sim 17\%/mW}$, which corresponds to a visible-to-telecom DFG photon (not power) conversion efficiency of $\mathrm{\sim 7.2\%/mW}$. The lower DFG value is due to the difference in initial and converted photon frequencies. The outlook for the triple resonance microring approach for large wavelength shift quantum frequency conversion is promising.
Absolute conversion efficiency can be further increased with higher quality factors. We have observed undercoupled quality factors as high as $\mathrm{\sim 3 \times 10^5}$ at both telecom and visible wavelengths for similar GaP-on-insulator devices (Supplement 1), suggesting a small-signal DFG photon conversion efficiency of $\mathrm{\sim 45\%/mW}$ is possible in the near term, with near-unity conversion in over-coupled devices with sufficient pump power.

Efficiency, however, is only one aspect in evaluating QFC performance. Because 637\,nm to telecom conversion requires a pump at a shorter wavelength than the output, telecom noise photons can potentially be generated by spontaneous down conversion, a known issue in PPLN devices~\cite{morrison2021bright}. There may, however, be an additional advantage to resonant devices which suppress undesired frequency conversion processes by a combination of phase matching and resonance. Optimal waveguide-resonator coupling and efficient off-chip coupling, while theoretically straightforward, pose engineering challenges. Perhaps the most critical challenge, however, will be achieving triple resonance from a precise defect optical transition frequency to a target telecom frequency. This will require further advancement in multiple independent, or quasi-independent resonator tuning mechanisms~\cite{thiel2022trimming,Lu2020splitting}. Once achieved, the impact of triply-resonant devices would extend to wavelength multiplexing for quantum networks and the elimination of emitter wavelength inhomogeneity.  

\section{Acknowledgement}
This material is based upon work supported by the National Science Foundation under grants EFMA-1640986 (photonic design, fabrication and passive device testing) and U.S. Department of Energy, Office of Science, National Quantum Information Science Research Centers, Co-design Center for Quantum Advantage (C2QA) under contract number DE-SC0012704 (frequency conversion testing). The photonic devices were fabricated at the Washington Nanofabrication Facility, a National Nanotechnology Coordinated Infrastructure (NNCI) site at the University of Washington which is supported in part by funds from the National Science Foundation (awards NNCI-2025489, 1542101, 1337840 and 0335765). We thank Mo Li for the use of optical test equipment.

\renewcommand{\thesection}{SI.\arabic{section}}
\renewcommand{\theequation}{SI.\arabic{equation}}
\renewcommand{\thefigure}{SI.\arabic{figure}}
\renewcommand{\thetable}{SI.\arabic{table}}
\setcounter{section}{0}

\section{Supplemental Information}

\subsection{Photonic integrated circuit design considerations}

Aperiodic grating couplers were designed to provide off-chip coupling at each wavelength with minimal footprint.
Grating designs for each wavelength were generated using a simplified model treating the grating as series of discrete, independent scattering elements in a waveguide.
Scatterer widths and spacings were optimized to maximize overlap between upward-scattered light and a target Gaussian mode shape using a sampled discrete hill climbing algorithm.
The detailed coupling behavior from waveguide to free space was simulated by Lumerical 3D FDTD for each grating design candidate.
The grating design that provided the highest coupling into a single spot in free space over the target wavelength band was selected.
If the simulated grating efficiency peak was offset from the target wavelength, all lengths in the grating design were scaled by up to 5\% to red- or blue-shift the efficiency curve.
For grating couplers intended for multi-wavelength operation, two design approaches were used.
First, gratings that had been designed for one wavelength were FDTD simulated at the second wavelength. If an existing design also provided reasonable performance in the second wavelength band, it would be used as a multi-wavelength coupler.
Alternatively, grating designs can be co-optimized for free space coupling at both wavelengths using the hill-climbing algorithm to maximize the product of the two single-wavelength objective functions. However, this second strategy was not employed in this work.
The measured grating coupler efficiencies at the telecom, NIR, and visible resonant wavelengths ranged between 11-20\%, 5.5-8.5\% and 1-2\% in these devices.
At the design wavelengths of 1550, 1080, and 637\,nm, the respective grating efficiencies were simulated as 20\%, 17\%, and 40\%, but the NIR and visible gratings were operated at non-optimal wavelengths.

Wrapped-waveguide ring resonator coupling regions were designed using Lumerical MODE bent-waveguide eigenmode simulations for an extended supermode analysis.
For different combinations of coupling waveguide width ($w_{wg}$) and ring-waveguide separation distance ($d$), mode profiles and propagation constants were simulated for all modes in the combined ring and waveguide structure.
For each combined-structure mode, mode overlaps were calculated for both the ring resonator mode and the targeted waveguide mode.
During a single pass through the coupling region, light in the ring mode is divided between the combined-structure modes based on the mode overlaps.
The relative phases of these modes evolve as the light propagates in the combined structure for some angular distance $\theta$ around the ring.
The mode phases affect how much light is coupled into the waveguide mode at the end of the coupling region, and how much remains in the ring.
The coupling quality factor $Q_c$ can then be calculated based on the single-pass ring-to-waveguide coupling and the round-trip propagation time of the ring mode.
Values of $w_{wg}$, $d$, and $\theta$ were selected to provide targeted values of $Q_c$, which were chosen in a range around experimentally observed quality factors in similar photonic devices \cite{logan2018shg}.

\begin{figure}[h!]
\centering\includegraphics[width=13cm]{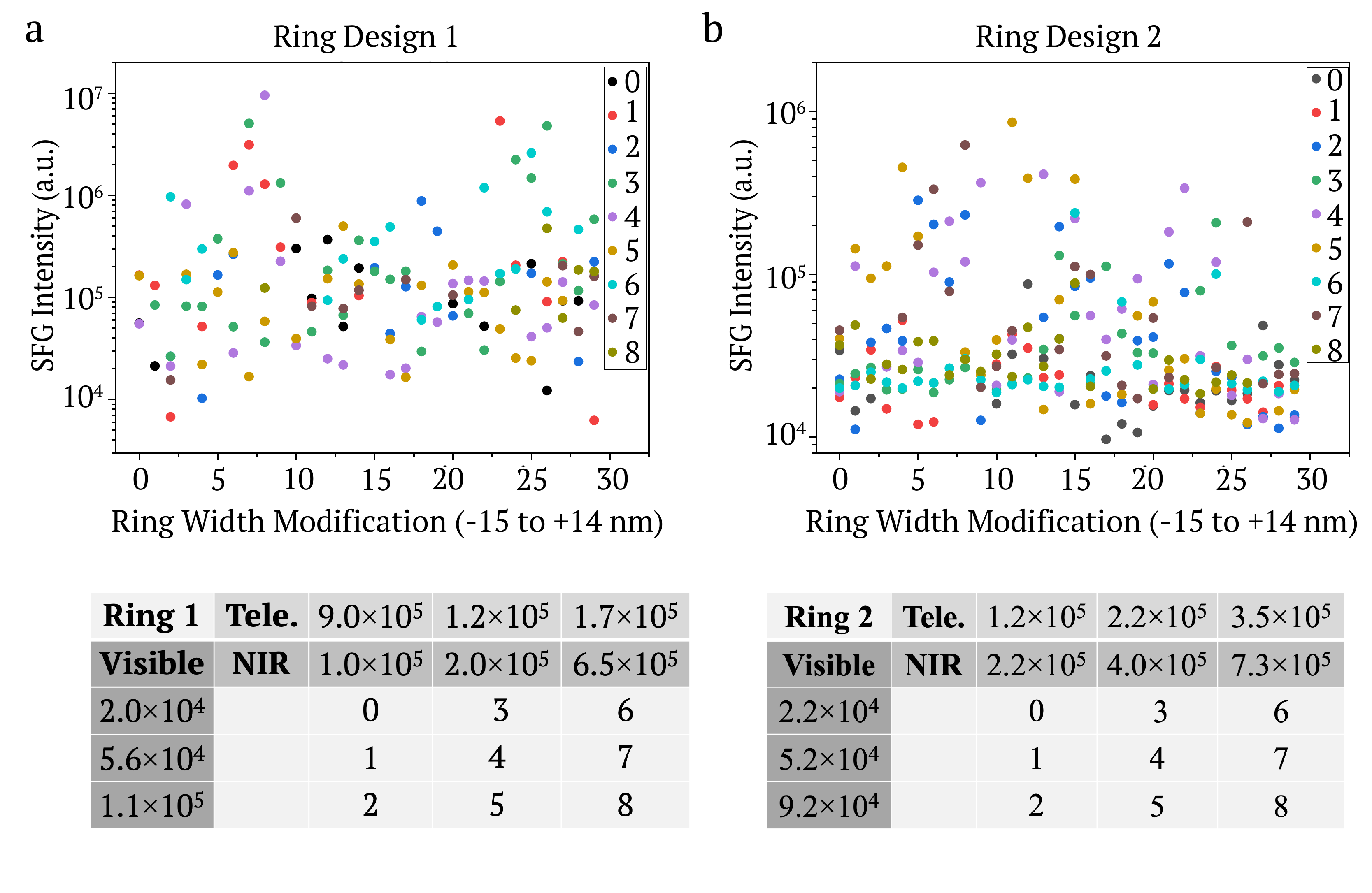}
\caption{Supercontinuum-pumped SFG performance survey showing peak SFG intensity at $\mathrm{24^{\circ}C}$ for all devices of ring designs 1 (a) and 2 (b). All devices were excited with similar input powers, so the measured SFG intensities approximately correspond to conversion efficiencies relative to other devices. While most of the devices are not at their optimal temperatures for SFG, particularly bright devices are likely to be within temperature tuning range ($\mathrm{14-40^{\circ}C}$) of triple resonance. Data points are colored according to the simulated coupling quality factors for each device, which are summarized for each ring design in the tables below.} 
\label{sup_fig_survey}
\end{figure}

Due to previous observations of pump-band ring modes coupling to telecom-band output waveguides and vice versa, both infrared ring modes were coupled to a single waveguide in these devices.
Wrapped-waveguide coupling regions were designed for the first ring design, which used fundamental radially-polarized (TE) modes at both wavelengths.
However, a wrapped waveguide design could not be found that could accommodate both the telecom mode and higher-order pump mode in the second ring design.
Instead, a straight-waveguide coupling region was used for the infrared modes for that ring design.
For straight waveguides, single-pass coupling was simulated using 3D FDTD, and coupling quality factor was calculated as normal.
The visible and infrared $Q_c$ targets used for each ring design can be found in Fig.~\ref{sup_fig_survey}.

The primary design variables for the ring resonators are ring width $w$ and radius $r$, since the GaP layer thickness is constant across the MBE chip.
The ring dimensions are optimized to maximize nonlinear overlap $\beta$ while satisfying the quasi-phase matching condition $M = \pm 2$ (Eqn.~3) for a selected set of telecom (1550\,nm), visible (637\,nm), and pump (1081.4\,nm) modes.
At each design point, propagation constants and mode profiles for each mode are simulated using Lumerical MODE bent-waveguide eigenmode solver.
The propagation constants are used to calculate azimuthal mode numbers $m_n$, which may be fractional if the precise design wavelengths are not resonant, and $M = m_1 + m_2 - m_3$.
Nonlinear overlap $\beta$ is calculated from the mode profiles.
To find valid ring designs, first an arbitrary combination of $r$ and $w$ is evaluated.
If  quasi-phase matching (QPM) is not satisfied, the ring width is iteratively modified until a valid design is found ($M \approx \pm 2$).
Ring dimensions and $\beta$ for this design are recorded. Then, ring radius is adjusted, and ring width is again tuned until the same QPM condition is satisfied.
This process is repeated while following increasing trends in $\beta$ (usually, toward smaller radius) until either $\beta$ is maximized or the ring width or radius becomes too small to confine the highest-wavelength mode without severe degradation of quality factor. 

\begin{figure}[ht]
\centering\includegraphics[width=13cm]{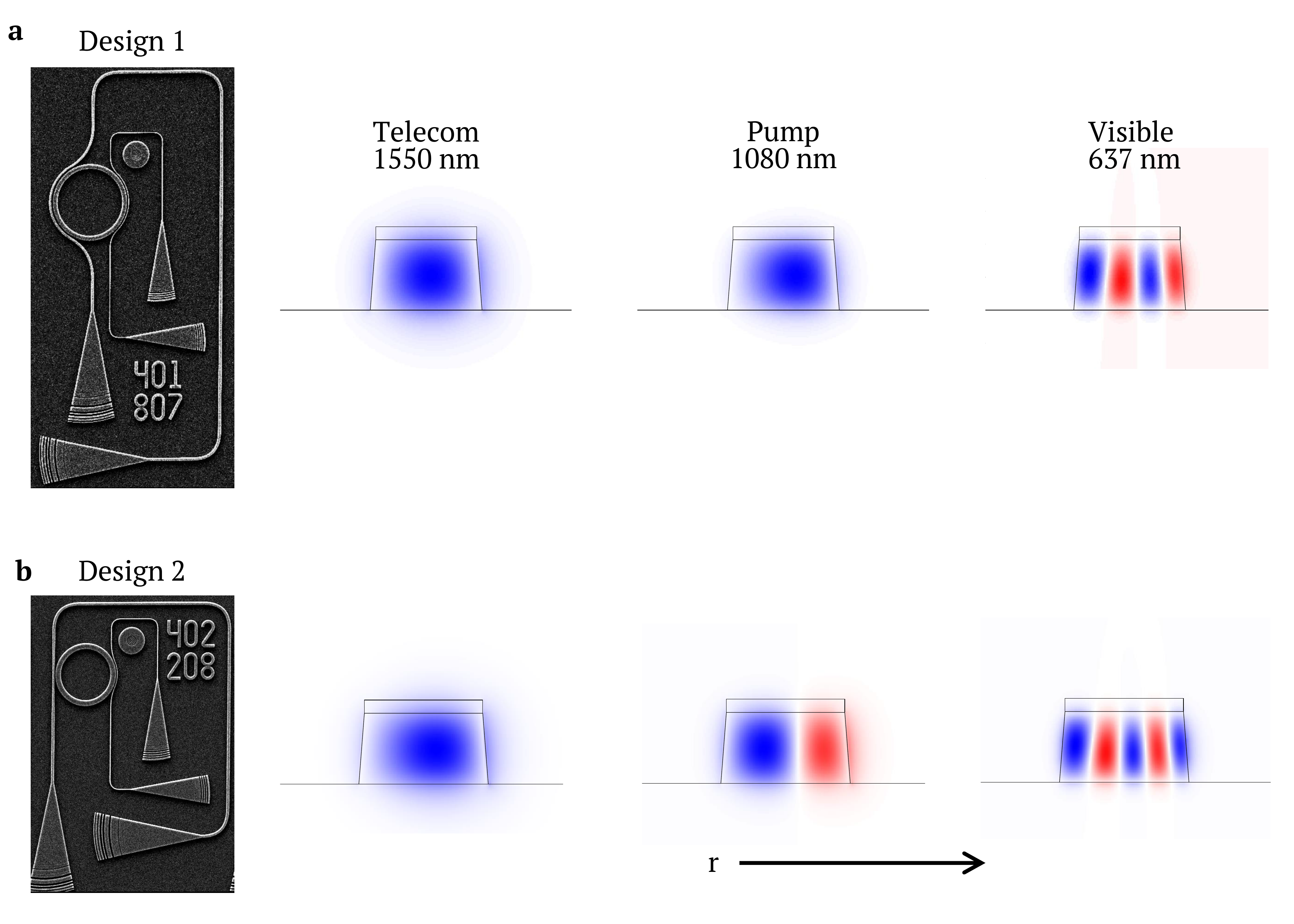}
\caption{SEM images and mode profiles for the primary (a) and alternative (b) ring designs fabricated for this work.} 
\label{sup_ring_design}
\end{figure}

The first ring design in this work was designed with the mode combination of $\mathrm{TE_{00}}$, $\mathrm{TE_{00}}$, and $\mathrm{TM_{03}}$, respectively, for the telecom, pump, and visible modes.
With a ring width of 690\,nm and radius of 5300\,nm, this mode combination was simulated to satisfy the QPM condition $M=-2$, with a nonlinear overlap of $|\beta|^2 = \mathrm{27.8\,J^{-1}}$.
As described in Eqn. 3, a ring resonator in a [1 0 0]-normal zincblende crystal requires two TE modes and one vertically-polarized (TM) mode to provide significant nonlinear overlap $\beta$. The shortest wavelength mode was selected to be the TM mode due to previous observations and simulations that TM mode quality factor is more sensitive to material thickness than TE modes.
Satisfying the QPM condition purely by mode engineering requires a higher-order visible mode, with the trade-off that $\beta$ is usually higher for modes that are closer to the fundamental.
The $\mathrm{TE_{00}}$/$\mathrm{TE_{00}}$/$\mathrm{TM_{03}}$ combination was selected because it allows phase matching with a ring width that is wide enough to easily accommodate the fundamental telecom mode. 
Ring designs with $\mathrm{TE_{00}}$/$\mathrm{TE_{00}}$/$\mathrm{TM_{02}}$ modes could provide higher values of $\beta$, but achieving quasi-phase matching with these modes requires substantially narrower rings with poor confinement of the telecom mode.
Conversely, $\mathrm{TE_{00}}$/$\mathrm{TE_{00}}$/$\mathrm{TM_{04}}$ allows for wider ring resonators at the cost of severe degradation of the nonlinear overlap.
A second ring design, shown in Fig.~\ref{sup_ring_design}, was developed using a $\mathrm{TE_{00}}$/$\mathrm{TE_{01}}$/$\mathrm{TM_{04}}$ telecom/pump/visible mode combination, which produced a higher nonlinear overlap $|\beta|^2 = \mathrm{68.\,J^{-1}}$ with a ring width of 810\,nm and radius of 4425\,nm.
However, due to poor coupling and reduced quality factor of the pump mode, devices fabricated with the second design exhibited substantially lower conversion efficiency, as shown in Fig.~\ref{sup_fig_survey}.

\subsection{Supercontinuum pump survey of all devices}
\begin{figure}[h!]
\centering\includegraphics[width=13cm]{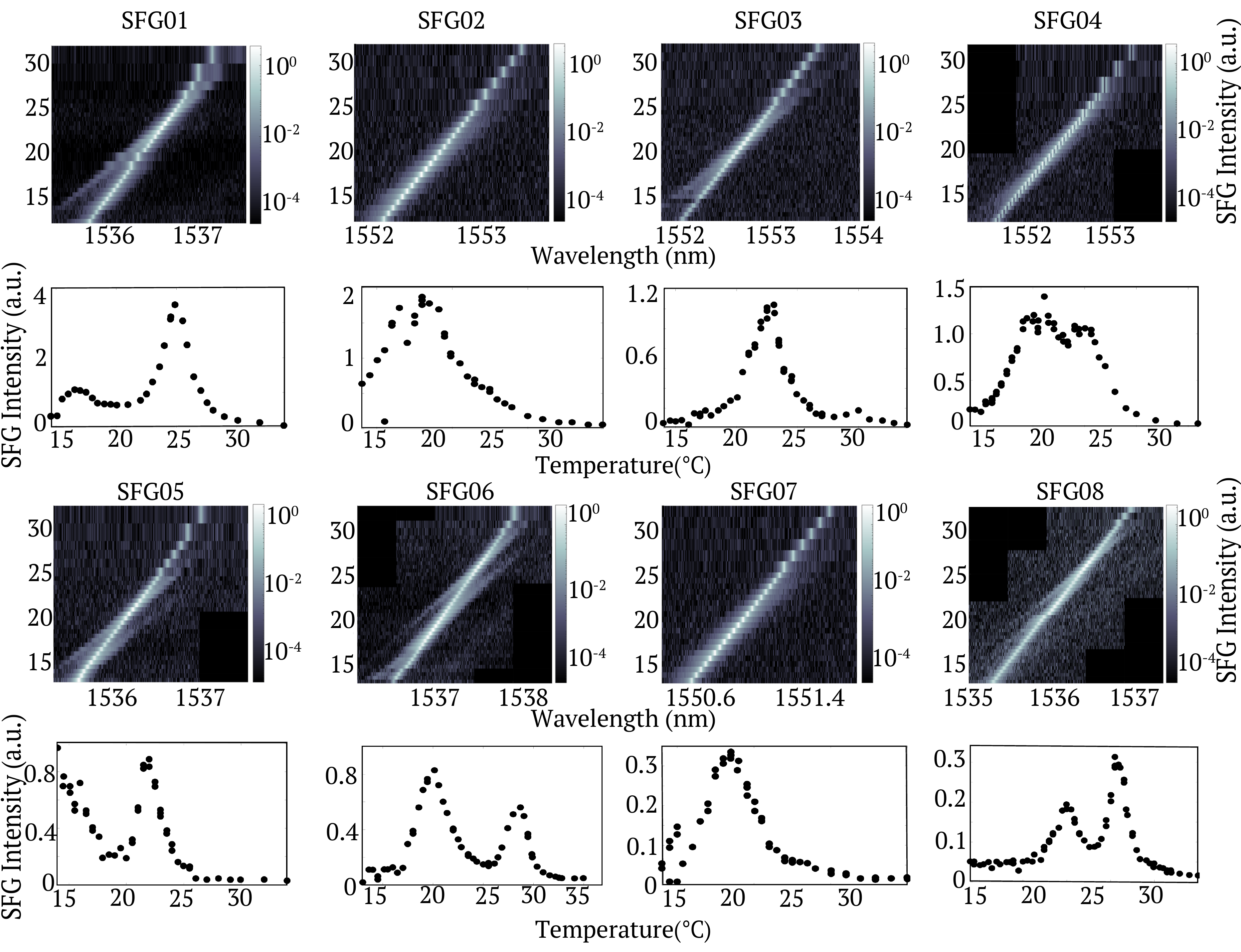}
\caption{Temperature dependence of supercontinuum-pump SFG intensity for devices SFG01-SFG08 (a-h). For each device, a logarithmic-scale color plot shows the SFG dependence on both temperature and telecom-band input wavelength. Every pump-band mode is always resonantly excited by some part of the filtered supercontinuum laser input (1000-1200\,nm). The primary intensity line follows the telecom-band resonance, while dimmer lines associated with visible-band resonances tune across the telecom resonance to produce peaks in the SFG conversion efficiency. A linear-scale plot of the peak SFG intensity at each temperature is also shown for each device.} 
\label{sup_fig_temp_survey}
\end{figure}

In order to compensate for variations in the fabrication process, arrays of devices of both ring designs were fabricated with varying ring widths and coupling region parameters.
The supercontinuum-pumped SFG method described in the main paper was used to identify candidate devices for more detailed frequency conversion testing.
First, a survey of peak SFG intensity at room temperature was performed on all operational ring resonators of both designs, shown in Fig.~\ref{sup_fig_survey}.
Because the brightest observed devices of the second design produced roughly 10\% of the SFG intensity of the first design rings, subsequent testing was focused on the first design.
The supercontinuum-pumped SFG survey was repeated for the design 1 devices at temperatures from $\mathrm{16-40^{\circ}C}$ to identify devices with the strong temperature dependence indicating a triply-resonant conversion process.
Eight devices with bright, temperature-sensitive SFG were selected for additional testing.
Supercontinuum-pumped SFG was measured at $\mathrm{0.5^{\circ}C}$ intervals to find the optimal temperature and corresponding peak SFG intensity for each of these devices, as shown in Fig.~\ref{sup_fig_temp_survey}.

\begin{figure}[h!]
\centering\includegraphics[width=13cm]{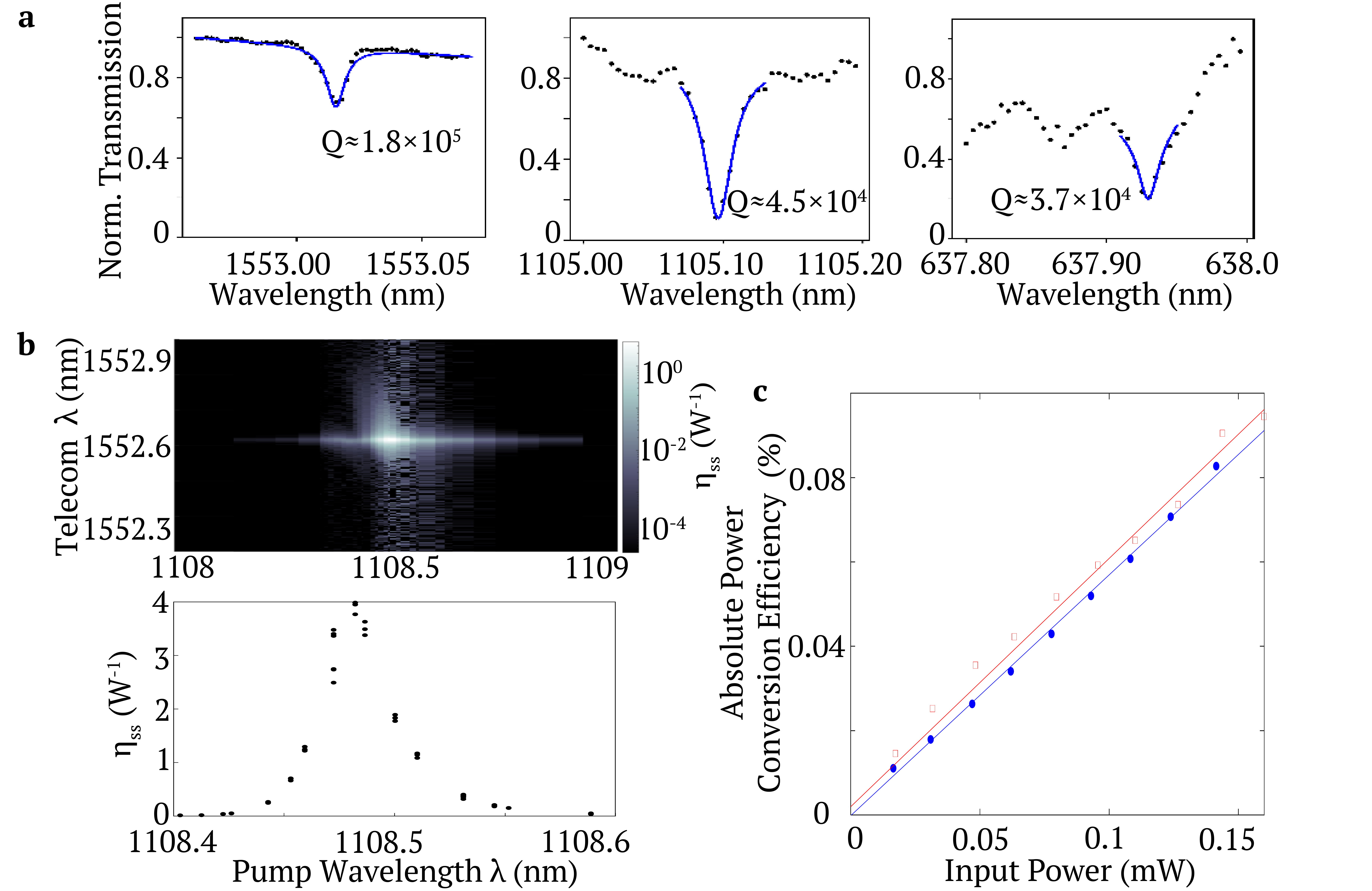}
\caption{(a) Transmission measurements of the telecom, near-infrared, and visible resonances in device SFG02, along with fitted Lorentzian curves and quality factors. Background is approximated with linear function. (b) Supercontinuum-pumped SFG intensity as a function of temperature. (c) Power conversion efficiency of device SFG02 as a function of telecom (blue) and NIR (red) input power. For each data series, the other wavelength input is constant and weak relative to the variable input power.} 
\label{sup_fig_dev2}
\end{figure}

The two devices that exhibited the highest peak SFG intensity, designated SFG01 and SFG02, were selected for calibrated conversion efficiency testing.
The SFG testing methods and performance for SFG01 are described in the main text.
As shown in Fig.~\ref{sup_fig_dev2}, device SHG02 was characterized using similar procedures, which found a small-signal power conversion efficiency of $\mathrm{370-580\%/W}$.
Due to mode splitting in this device, converted light was observed coupling out of both visible-band grating couplers. The conversion efficiency was calculated from the output power of a single grating coupler.

For both devices, power conversion efficiency is calculated as $\eta = P_{\text{out}} / P_{\text{in}}$, where $P_{\text{out}}$ is the visible SFG power in the output waveguide and $P_{\text{in}}$ is either the telecom or NIR power in the input waveguide, depending which wavelength is designated as the pump in each measurement.
In either case, the small-signal conversion efficiency $\eta_{\text{ss}} = P_{\text{vis}}/P_{\text{telecom}}P_{\text{NIR}}$ is the slope of the power dependence of $\eta$ at low input powers.
Photon conversion efficiency is power conversion efficiency scaled by the ratio of input and output photon energies.
For each wavelength, power in the waveguide is calculated from off-chip power measurements combined with the grating coupler efficiencies for the device under test.
For the telecom and NIR band inputs, the input transmission efficiencies from the point of measurement to the grating coupler were 56.5\% and 77.2\%, respectively.
The total collection and detection efficiency of the visible-band output at 647\,nm was 2.4\%.

\begin{figure}[ht]
\centering\includegraphics[width=13cm]{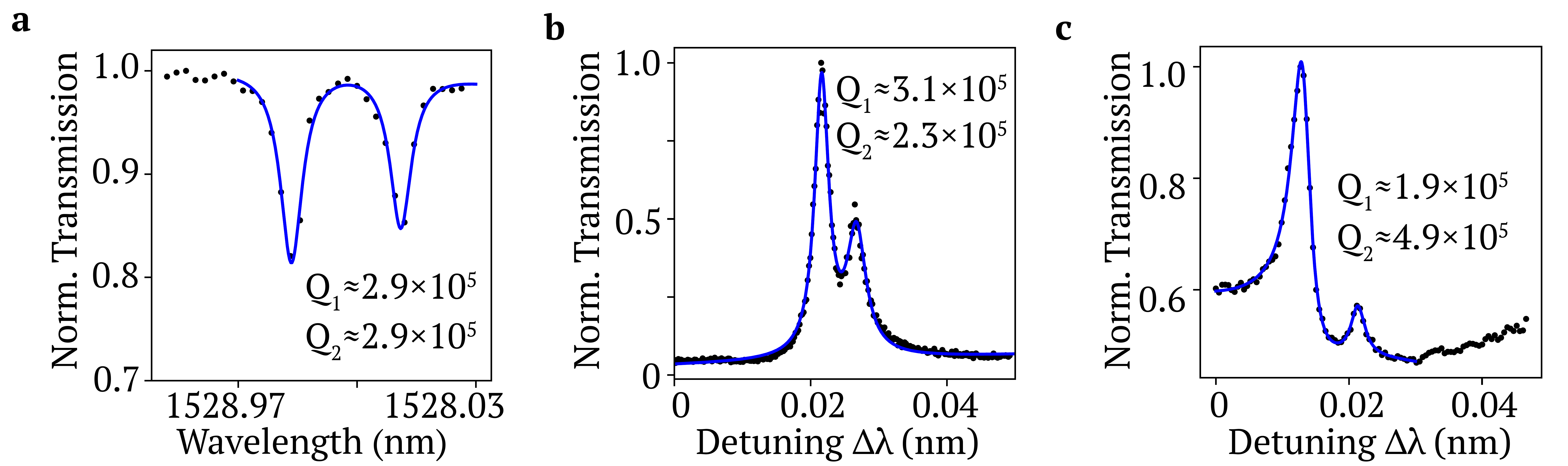}
\caption{(a) Transmission measurements of the telecom resonances in a SFG device, along with fitted Lorentzian curves and quality factors. Background is approximated with linear function. 
(b--c) Drop port transmission measurements of TM-mode visible resonances in a 5-{\textmu}m diameter GaP disk resonator on silicon nitride, obtained by scanning a tunable laser. The positions of each resonance are determined to be (b) $768.38\pm 0.16$ nm and (c) $635.05\pm 0.11$ nm respectively, with Lorentzian fits giving quality factors as large as half million. Here both resonances are in the undercoupled regime demonstrating the large intrinsic quality factor in the visible band. The PL lineshape is fitted using a Fano and a  Lorentzian function.} 
\label{Transmission-fit}
\end{figure}

\subsection{Highest quality factors measured in GaP-on-insulator devices}

The GaP-on-insulator platform has the potential for significantly higher performance if higher Q-factors can be obtained. The Q-factors in the highest performing SFG devices in this study are lower that what has been obtained in other devices. We have observed telecom Q's as high as $290,000$ on the SFG chip (Fig. SI.5a). We have measured visible Q's ranging from 200,000-490,000 in 290-nm thick GaP disk resonators on nitride (Fig. SI.5b,c). These quality factors are all measured in the undercoupled regime. At critical coupling, Q's may reduce by up to a factor of two. For the visible transmission measurements, an add-drop port geometry is utilized with the drop port monitored.


\begin{thebibliography}{10}
\newcommand{\enquote}[1]{``#1''}

\bibitem{kumar1990qfc}
P.~Kumar, \enquote{Quantum frequency conversion,} {\protect\JournalTitle{Opt.
  Lett.}} \textbf{15}, 1476--1478 (1990).

\bibitem{bock2018high}
M.~Bock, P.~Eich, S.~Kucera, M.~Kreis, A.~Lenhard, C.~Becher, and J.~Eschner,
  \enquote{High-fidelity entanglement between a trapped ion and a telecom
  photon via quantum frequency conversion,} {\protect\JournalTitle{Nature
  communications}} \textbf{9}, 1--7 (2018).

\bibitem{van2020long}
T.~van Leent, M.~Bock, R.~Garthoff, K.~Redeker, W.~Zhang, T.~Bauer,
  W.~Rosenfeld, C.~Becher, and H.~Weinfurter, \enquote{Long-distance
  distribution of atom-photon entanglement at telecom wavelength,}
  {\protect\JournalTitle{Physical review letters}} \textbf{124}, 010510 (2020).

\bibitem{zaske2012visible}
S.~Zaske, A.~Lenhard, C.~A. Ke{\ss}ler, J.~Kettler, C.~Hepp, C.~Arend,
  R.~Albrecht, W.-M. Schulz, M.~Jetter, P.~Michler \emph{et~al.},
  \enquote{Visible-to-telecom quantum frequency conversion of light from a
  single quantum emitter,} {\protect\JournalTitle{Physical review letters}}
  \textbf{109}, 147404 (2012).

\bibitem{tchebotareva2019entanglement}
A.~Tchebotareva, S.~L.~N. Hermans, P.~C. Humphreys, D.~Voigt, P.~J. Harmsma,
  L.~K. Cheng, A.~L. Verlaan, N.~Dijkhuizen, W.~de~Jong, A.~Dr\'eau, and
  R.~Hanson, \enquote{Entanglement between a diamond spin qubit and a photonic
  time-bin qubit at telecom wavelength,} {\protect\JournalTitle{Phys. Rev.
  Lett.}} \textbf{123}, 063601 (2019).

\bibitem{walker2018long}
T.~Walker, K.~Miyanishi, R.~Ikuta, H.~Takahashi, S.~V. Kashanian, Y.~Tsujimoto,
  K.~Hayasaka, T.~Yamamoto, N.~Imoto, and M.~Keller, \enquote{Long-distance
  single photon transmission from a trapped ion via quantum frequency
  conversion,} {\protect\JournalTitle{Physical review letters}} \textbf{120},
  203601 (2018).

\bibitem{dreau2018nvqfc}
A.~Dr\'eau, A.~Tchebotareva, A.~E. Mahdaoui, C.~Bonato, and R.~Hanson,
  \enquote{Quantum frequency conversion of single photons from a
  nitrogen-vacancy center in diamond to telecommunication wavelengths,}
  {\protect\JournalTitle{Phys. Rev. Applied}} \textbf{9}, 064031 (2018).

\bibitem{ates2012two}
S.~Ates, I.~Agha, A.~Gulinatti, I.~Rech, M.~T. Rakher, A.~Badolato, and
  K.~Srinivasan, \enquote{Two-photon interference using background-free quantum
  frequency conversion of single photons emitted by an inas quantum dot,}
  {\protect\JournalTitle{Physical review letters}} \textbf{109}, 147405 (2012).

\bibitem{weber2019twophoton}
J.~H. Weber, B.~Kambs, J.~Kettler, S.~Kern, J.~Maisch, H.~Vural, M.~Jetter,
  S.~L. Portalupi, C.~Becher, and P.~Michler, \enquote{Two-photon interference
  in the telecom c-band after frequency conversion of photons from remote
  quantum emitters,} {\protect\JournalTitle{Nature Nanotechnology}}
  \textbf{14}, 23--26 (2019).

\bibitem{levonian2022distinguishable}
D.~S. Levonian, R.~Riedinger, B.~Machielse, E.~N. Knall, M.~K. Bhaskar, C.~M.
  Knaut, R.~Bekenstein, H.~Park, M.~Lon\ifmmode~\check{c}\else \v{c}\fi{}ar,
  and M.~D. Lukin, \enquote{Optical entanglement of distinguishable quantum
  emitters,} {\protect\JournalTitle{Phys. Rev. Lett.}} \textbf{128}, 213602
  (2022).

\bibitem{albota2004bulkqfc}
M.~A. Albota and F.~N. Wong, \enquote{Efficient single-photon counting at 1.55
  $\mu$m by means of frequency upconversion,} {\protect\JournalTitle{Optics
  letters}} \textbf{29}, 1449--1451 (2004).

\bibitem{rutz2017quantum}
H.~R{\"u}tz, K.-H. Luo, H.~Suche, and C.~Silberhorn, \enquote{Quantum frequency
  conversion between infrared and ultraviolet,} {\protect\JournalTitle{Physical
  Review Applied}} \textbf{7}, 024021 (2017).

\bibitem{allgaier2017conversion}
M.~Allgaier, V.~Ansari, L.~Sansoni, C.~Eigner, V.~Quiring, R.~Ricken,
  G.~Harder, B.~Brecht, and C.~Silberhorn, \enquote{Highly efficient frequency
  conversion with bandwidth compression of quantum light,}
  {\protect\JournalTitle{Nature Communications}} \textbf{8}, 14288 (2017).

\bibitem{samblowski2014conversion}
A.~Samblowski, C.~E. Vollmer, C.~Baune, J.~Fiur\'{a}\v{s}ek, and R.~Schnabel,
  \enquote{Weak-signal conversion from 1550 to 532 nm with 84\% efficiency,}
  {\protect\JournalTitle{Opt. Lett.}} \textbf{39}, 2979--2981 (2014).

\bibitem{guerreiro2013interaction}
T.~Guerreiro, E.~Pomarico, B.~Sanguinetti, N.~Sangouard, J.~Pelc, C.~Langrock,
  M.~Fejer, H.~Zbinden, R.~T. Thew, and N.~Gisin, \enquote{Interaction of
  independent single photons based on integrated nonlinear optics,}
  {\protect\JournalTitle{Nature communications}} \textbf{4}, 1--5 (2013).

\bibitem{pelc2011long}
J.~S. Pelc, L.~Ma, C.~R. Phillips, Q.~Zhang, C.~Langrock, O.~Slattery, X.~Tang,
  and M.~M. Fejer, \enquote{Long-wavelength-pumped upconversion single-photon
  detector at 1550 nm: performance and noise analysis,}
  {\protect\JournalTitle{Opt. Express}} \textbf{19}, 21445--21456 (2011).

\bibitem{stolk2022telecom}
A.~Stolk, K.~L. van~der Enden, M.-C. Roehsner, A.~Teepe, S.~O. Faes, S.~Cadot,
  J.~van Rantwijk, I.~t. Raa, R.~Hagen, A.~Verlaan \emph{et~al.},
  \enquote{Telecom-band quantum interference of frequency-converted photons
  from remote detuned nv centers,} {\protect\JournalTitle{arXiv preprint
  arXiv:2202.00036}}  (2022).

\bibitem{wang2021degenerate}
J.-Q. Wang, Y.-H. Yang, M.~Li, X.-X. Hu, J.~B. Surya, X.-B. Xu, C.-H. Dong,
  G.-C. Guo, H.~X. Tang, and C.-L. Zou, \enquote{Efficient frequency conversion
  in a degenerate ${\ensuremath{\chi}}^{(2)}$ microresonator,}
  {\protect\JournalTitle{Phys. Rev. Lett.}} \textbf{126}, 133601 (2021).

\bibitem{ye2020triple}
X.~Ye, S.~Liu, Y.~Chen, Y.~Zheng, and X.~Chen, \enquote{Sum-frequency
  generation in lithium-niobate-on-insulator microdisk via modal phase
  matching,} {\protect\JournalTitle{Opt. Lett.}} \textbf{45}, 523--526 (2020).

\bibitem{dreau2018quantum}
A.~Dr{\'e}au, A.~Tchebotareva, A.~El~Mahdaoui, C.~Bonato, and R.~Hanson,
  \enquote{Quantum frequency conversion of single photons from a
  nitrogen-vacancy center in diamond to telecommunication wavelengths,}
  {\protect\JournalTitle{Physical review applied}} \textbf{9}, 064031 (2018).

\bibitem{ikuta2014frequency}
R.~Ikuta, T.~Kobayashi, S.~Yasui, S.~Miki, T.~Yamashita, H.~Terai, M.~Fujiwara,
  T.~Yamamoto, M.~Koashi, M.~Sasaki \emph{et~al.}, \enquote{Frequency
  down-conversion of 637 nm light to the telecommunication band for
  non-classical light emitted from nv centers in diamond,}
  {\protect\JournalTitle{Optics express}} \textbf{22}, 11205--11214 (2014).

\bibitem{huang2021siv}
D.~Huang, A.~Abulnaga, S.~Welinski, M.~Raha, J.~D. Thompson, and N.~P. de~Leon,
  \enquote{Hybrid {III-V} diamond photonic platform for quantum nodes based on
  neutral silicon vacancy centers in diamond,} {\protect\JournalTitle{Opt.
  Express}} \textbf{29}, 9174--9189 (2021).

\bibitem{bersin2021siv}
E.~Bersin, N.~Wan, M.~Bhaskar, D.~Levonian, R.~Riedinger, C.~Langrock, M.~M.
  Fejer, M.~Lukin, P.~B. Dixon, S.~Hamilton, and D.~Englund, \enquote{A
  low-noise telecom interface for silicon-vacancy quantum network nodes,} in
  \emph{2021 Conference on Lasers and Electro-Optics (CLEO),}  (2021), pp.
  1--2.

\bibitem{bond1965measurement}
W.~Bond, \enquote{Measurement of the refractive indices of several crystals,}
  {\protect\JournalTitle{Journal of Applied Physics}} \textbf{36}, 1674--1677
  (1965).

\bibitem{dal1996density}
A.~Dal~Corso, F.~Mauri, and A.~Rubio, \enquote{Density-functional theory of the
  nonlinear optical susceptibility: Application to cubic semiconductors,}
  {\protect\JournalTitle{Physical Review B}} \textbf{53}, 15638 (1996).

\bibitem{rivoire2011second}
K.~Rivoire, S.~Buckley, F.~Hatami, and J.~Vu{\v{c}}kovi{\'c}, \enquote{Second
  harmonic generation in {GaP} photonic crystal waveguides,}
  {\protect\JournalTitle{Applied Physics Letters}} \textbf{98}, 263113 (2011).

\bibitem{barclay2009hybrid}
P.~E. Barclay, K.-M. Fu, C.~Santori, and R.~G. Beausoleil, \enquote{Hybrid
  photonic crystal cavity and waveguide for coupling to diamond nv-centers,}
  {\protect\JournalTitle{Optics Express}} \textbf{17}, 9588--9601 (2009).

\bibitem{schneider2018gallium}
K.~Schneider, P.~Welter, Y.~Baumgartner, H.~Hahn, L.~Czornomaz, and P.~Seidler,
  \enquote{Gallium phosphide-on-silicon dioxide photonic devices,}
  {\protect\JournalTitle{Journal of Lightwave Technology}} \textbf{36},
  2994--3002 (2018).

\bibitem{englund2010deterministic}
D.~Englund, B.~Shields, K.~Rivoire, F.~Hatami, J.~Vuckovic, H.~Park, and M.~D.
  Lukin, \enquote{Deterministic coupling of a single nitrogen vacancy center to
  a photonic crystal cavity,} {\protect\JournalTitle{Nano letters}}
  \textbf{10}, 3922--3926 (2010).

\bibitem{burgess2009difference}
I.~B. Burgess, A.~W. Rodriguez, M.~W. McCutcheon, J.~Bravo-Abad, Y.~Zhang,
  S.~G. Johnson, and M.~Lon{\v{c}}ar, \enquote{Difference-frequency generation
  with quantum-limited efficiency in triply-resonant nonlinear cavities,}
  {\protect\JournalTitle{Optics express}} \textbf{17}, 9241--9251 (2009).

\bibitem{thiel2022trimming}
L.~Thiel, A.~D. Logan, S.~Chakravarthi, S.~Shree, K.~Hestroffer, F.~Hatami, and
  K.-M.~C. Fu, \enquote{Precise electron beam-based target-wavelength trimming
  for frequency conversion in integrated photonic resonators,}
  {\protect\JournalTitle{Opt. Express}} \textbf{30}, 6921--6933 (2022).

\bibitem{logan2018shg}
A.~D. Logan, M.~Gould, E.~R. Schmidgall, K.~Hestroffer, Z.~Lin, W.~Jin,
  A.~Majumdar, F.~Hatami, A.~W. Rodriguez, and K.-M.~C. Fu, \enquote{{400\%/W}
  second harmonic conversion efficiency in 14 {\textmu}m-diameter gallium
  phosphide-on-oxide resonators,} {\protect\JournalTitle{Opt. Express}}
  \textbf{26}, 33687--33699 (2018).

\bibitem{morrison2021bright}
C.~L. Morrison, M.~Rambach, Z.~X. Koong, F.~Graffitti, F.~Thorburn, A.~K. Kar,
  Y.~Ma, S.-I. Park, J.~D. Song, N.~G. Stoltz, D.~Bouwmeester, A.~Fedrizzi, and
  B.~D. Gerardot, \enquote{A bright source of telecom single photons based on
  quantum frequency conversion,} {\protect\JournalTitle{Applied Physics
  Letters}} \textbf{118}, 174003 (2021).

\bibitem{Lu2020splitting}
X.~Lu, A.~Rao, G.~Moille, D.~A. Westly, and K.~Srinivasan, \enquote{Universal
  frequency engineering tool for microcavity nonlinear optics: multiple
  selective mode splitting of whispering-gallery resonances,}
  {\protect\JournalTitle{Photon. Res.}} \textbf{8}, 1676--1686 (2020).

\end{thebibliography}
\end{document}